\documentclass[fleqn,usenatbib]{mnras}

\usepackage{newtxtext,newtxmath}

\usepackage[T1]{fontenc}
\usepackage{ae,aecompl}

\usepackage{graphicx}	
\usepackage{amsmath}	
\usepackage{amssymb}	
\usepackage{xcolor}

\title[Multiple Populations in LMC GCs]{Exploring the nature and synchronicity of early cluster formation in the Large Magellanic Cloud V: Multiple Populations in ancient Globular Clusters}

\author[Gilligan et al.]{
Christina K. Gilligan,$^{1}$\thanks{E-mail: christina.k.gilligan.gr@dartmouth.edu}
Brian Chaboyer,$^{1}$
Jeffrey D. Cummings,$^{2}$
Dougal Mackey,$^{3}$
\newauthor
Roger E. Cohen,$^{4}$
Douglas Geisler,$^{5}$ $^{6}$ $^{7}$
Aaron J. Grocholski,$^{8}$
M. C. Parisi,$^{9}$ 
\newauthor
Ata Sarajedini,$^{10}$
Paolo Ventura,$^{11}$
Sandro Villanova,$^{5}$
Soung-Chul Yang,$^{12}$
\newauthor
and Rachel Wagner-Kaiser$^{13}$
\\
$^{1}$Department of Physics and Astronomy, Dartmouth College, Hanover, NH 03784, USA\\
$^{2}$Center for Astrophysical Sciences, Johns Hopkins University, 3400 N. Charles Street, Baltimore, MD 21218, USA\\
$^{3}$Research School of Astronomy \& Astrophysics, Australian National University, Canberra, ACT 2611, Australia\\
$^{4}$Space Telescope Science Institute, Baltimore, MD 21218, USA \\
$^{5}$Departamento de Astronom\' ia, Casilla 160-C ,Universidad de Concepci\' on, Chile \\
$^{6}$Instituto de Investigaci\' on Multidisciplinario en Ciencia y Tecnolog\' ia, Universidad de La Serena, Avenida Ra\' ul Bitr\' an S/N, La Serena, Chile \\
$^{7}$Departamento de Astronom\' ia, Facultad de Ciencias, Universidad de La Serena, Av. Juan Cisternas 1200, La Serena, Chile \\ 
$^{8}$Department of Physics, American University, Washington, DC 20016, USA \\
$^{9}$Observatorio Astron\' omico, Universidad Nacional de C\' ordoba, Laprida 854, C\' ordoba, CP 5000, Argentina \\
$^{10}$Florida Atlantic University, 777 Glades Rd, SE-43, Room 256, Boca Raton, FL 33431, USA \\
$^{11}$INAF-Osservatorio Astronomico di Roma, Via Frascati 33, 00078 Monteporzio Catone (Roma), Italy \\ 
$^{12}$Korea Astronomy and Space Science Institute (KASI), Daejeon 305-348, Korea \\
$^{13}$Department of Astronomy, University of Florida, 211 Bryant Space Science Center, Gainesville, FL 32611, USA
}

\date{Accepted XXX. Received YYY; in original form ZZZ}

\pubyear{2015}

\hypersetup{draft}
\begin{document}
\label{firstpage}
\pagerange{\pageref{firstpage}--\pageref{lastpage}}
\maketitle

\begin{abstract}
We examine four ancient Large Magellanic Cloud (LMC) globular clusters (GCs) for evidence of multiple stellar populations using the Advanced Camera for Surveys and Wide Field Camera 3 on the Hubble Space Telescope Programme GO-14164. NGC 1466, NGC 1841, and NGC 2257 all show evidence for a redder, secondary population along the main-sequence. Reticulum does not show evidence for the presence of a redder population, but this GC has the least number of stars and Monte Carlo simulations indicate that the sample of main sequence stars is too small to robustly infer whether a redder population exists in this cluster. The second, redder, population of the other three clusters constitutes $\sim30-40\%$ of the total population along the main-sequence. This brings the total number of ancient LMC GCs with known split or broadened main-sequences to five. However, unlike for Hodge 11 and NGC 2210 (see \citet{Me}), none of the clusters show evidence for multiple populations in the horizontal branch. We also do not find evidence of a second population along the Red Giant Branch (RGB).
\end{abstract}




\section{Introduction} \label{sec:intro}

Globular clusters (GCs) are one of the most important objects to study in order to understand galaxy formation and evolution.  Many GCs are among the oldest objects in their host galaxies and are therefore more metal-poor than their host galaxy. However, we still do not completely understand the formation of GCs, with the most important open question being the reason for their multiple stellar populations.

Nearly all Galactic GCs that have been studied have shown evidence for multiple stellar populations \citep[e.g.][]{Piotto,Anderson,NGC6397}. In addition, many old GCs in the Large Magellanic Cloud (LMC) \citep[e.g.][]{Me,Milone2016,Milone2017,Mucciarelli2009} and Small Magellanic Cloud (SMC) \citep[e.g.][]{Paper1, Paper2, Paper3} have also been shown to have multiple stellar populations. 

There is both spectroscopic and photometric evidence for these multiple populations in GCs. The populations are photometrically discrete, with two or more separate populations visible in a colour-magnitude diagram (CMD) with the appropriate filters, rather than just a broadening of the CMD as would be seen with a population continuum.  

However, the light element abundances of these stars tend to have broad and overlapping abundance patterns rather than exhibiting in discrete groups. Performing spectroscopy of main-sequence stars in globular clusters is difficult even in the MW let alone other galaxies including the LMC and the SMC. Therefore, spectra of red giant branch (RGB) stars, among the brightest in the cluster, are commonly used to determine accurate abundances in GCs \citep[e.g.][]{Carretta, Nataf, Munoz, NGC5824, CarrettaNaO}. Nearly every cluster examined has a Na-O anticorrelation and using this anticorrelation each cluster can be divided into multiple stellar populations. Some clusters have an Mg-Al anticorrelation in addition. Interestingly, there are rarely iron abundance spreads \citep{Carretta}. However, there are rare clusters that do exhibit an [Fe/H] dispersion.

Elemental abundances can also be inferred using photometry, as long as the chosen filters are sensitive to these abundance differences. For example, \citet{Marino} shows that chromosome maps created from four different HST filters are sensitive to CNO, Fe and He abundances. They also find that in each GC even the primoridal population stars do not necessarily have homogeneous elemental abundances, in contrast to what has been found spectroscopically. 

One of the strongest pieces of evidence for multiple populations are split main-sequences (MS). Split MS have been commonly found in many massive Milky Way GCs \citep[e.g.][]{Piotto, Anderson, NGC6397} and in some LMC and SMC GCs \citep[e.g.][]{Milone2016, Milone2017, Paper1, Paper2, Paper3}. Studies of the SMC (\citet{Paper1, Paper2, Paper3}) found that all intermediate age and old GCs exhibited a split RGB. The only clusters in their sample which do not display split RGBs are the youngest, those younger than $\sim$2 Gyr. Their results indicate that massive clusters younger than $\sim$2 Gyr do not have split RGBs. \citet{Martocchia} found similar results with six LMC GCs of varying ages up to 11 Gyr.


However, even with all the advances in detecting the presence of multiple populations in GCs, their origin is still an open question. Both \citet{Bastian} and \citet{Renzini} give excellent overviews of the current theories of multiple population formation and their drawbacks and limitations. The four current competing formation scenarios of the multiple populations in GCs are: fast rotating massive stars, massive interacting binaries, supermassive stars, and AGB stars. Matching all of the available photometric and spectroscopic data with the theories has proven difficult, and presently none of the competing theories is able to account for all the observations satisfactorily. Clearly more data is needed to untangle the true cause of these multiple populations. By examining GCs in many different environments and of different ages, we can narrow down the cause of these multiple populations.

\citet{Me} was the first photometric study to look for multiple populations in ancient ($\sim$ 13 Gyr) LMC GCs. In \citet{Me}, we found that both NGC 2210 and Hodge 11 show strong evidence for multiple populations on the main-sequence and on the horizontal branch (HB) of Hodge 11. This current work examines the four other ancient LMC GCs from the same data set as \citet{Me}. NGC 2257, examined here, has been shown to have RGB stars of differing [O/Fe] and [Na/Fe] from spectroscopic analysis \citep{Mucciarelli2009}.

\section{Observations and Data Reduction}

Of the 15 known LMC GCs that are comparable in age to the oldest Galactic GCs, i.e. 13 Gyr \citep{OldLMC}, we observed 6 that were far from the bar of the LMC. This was to ensure that we are able to achieve high-quality photometry that is not greatly affected by field contamination and crowding in order to measure faint main-sequence stars, which are important to resolve multiple populations. Figure 1 of \citet{Rachel} shows the positions of each of these GCs in relation to the LMC. This paper analyzes 4 of the clusters: NGC 1466, NGC 1841, NGC 2257, and Reticulum with metallicities of [Fe/H] = -1.9, -2.02, -1.95, -1.57 respectively (\citet{Walker}, \citet{Grocholski}, \citet{Mucciarelli2010}, \citet{Grocholski}).

The data were taken with 54 orbits of the Hubble Space Telescope (HST) (GO-14164) in three broadband filters, F336W, F606W, and F814W. The F336W filter is sensitive to NH and CN lines, but F606W and F814W are largely insensitive to the light element abundance variations that are the primary indication of multiple populations. Therefore, we expect to see evidence for multiple populations in the F336W-F606W and F336W-F814W colours but not in the F606W-F814W colour. While the frequently used `magic trio' of filters (F275W, F336W, and F438W) are ideal to disentangle multiple populations due to CNO variations \citep{Piotto}, the filters that were chosen for this work only require one third the number of orbits that the `magic trio' would require to achieve the same S/N of the photometry. However, this choice of filters does reduce our sensitivity to the presence of multiple populations. Using F275W would be ideal, but due to the increased number of orbits required is not used for this set of observations. With the F336W filter, we are able to detect stars that are up to 5 magnitudes fainter than the main-sequence turnoff (MSTO). The data are reduced using Dolphot \citep{Dolphin}. A more complete description of the photometric pipeline is presented in \citet{Mackeyprep}. 




\section{Data Analysis}\label{Data}

The data analysis of these four GCs is performed in a similar manner to the procedure outlined in \citet{Me}, and we refer the interested reader to this paper for further details. Artificial star tests are used to estimate the photometric uncertainties of our sources since Dolphot underestimates the true uncertainties of our photometry due to crowding. This is discussed in greater detail in \citet{Mackeyprep}. 

After our data are reduced, we use a data cleansing pipeline to retrieve the most precise photometry possible. We clean our data by removing stars with large photometric uncertainties. However, if we set a given universal uncertainty cut-off, it would preferentially remove fainter stars since fainter stars inherently have larger errors. We divide our stars into bins based on magnitude and set the cut-off to 4 times the median uncertainty in each bin. Varying the binning parameters and the multiplicative factor of our uncertainty cut-off has no significant effect on our main results.

\begin{figure*}
  \centering
  \begin{tabular}{ccc}
    \includegraphics[width=0.7\columnwidth]{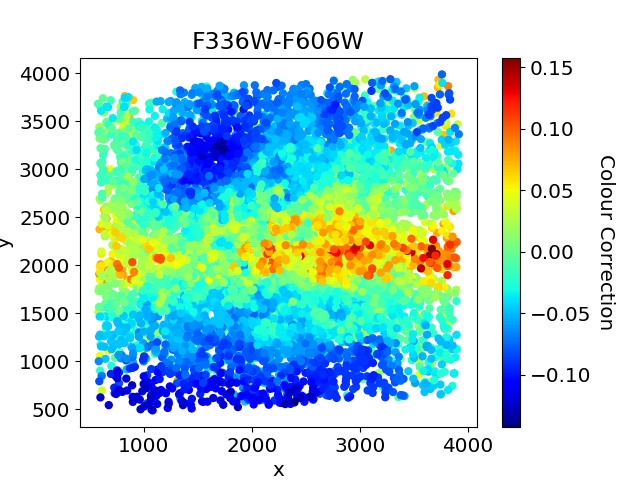} &
        \includegraphics[width=0.7\columnwidth]{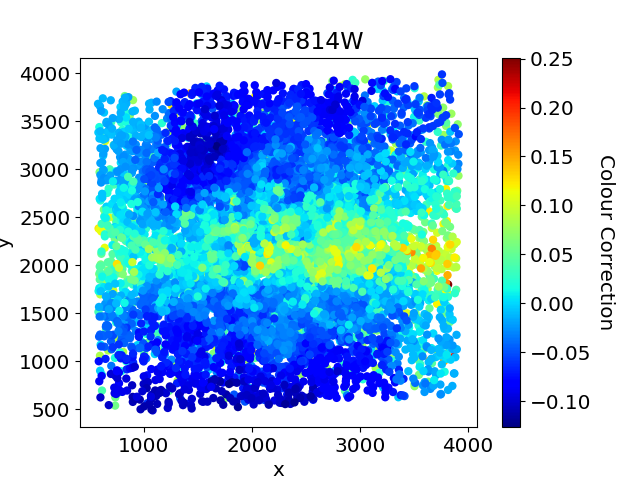} & \includegraphics[width=0.7\columnwidth]{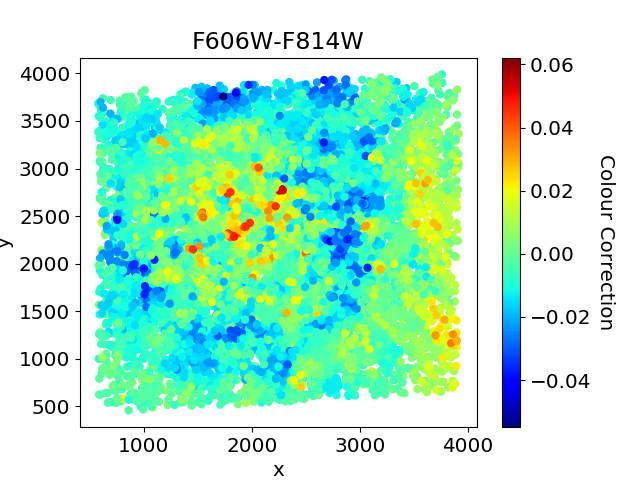}\\
        \small (a) &
                        \small (b)  & \small (c)\\
  \end{tabular}
  \caption{Colour correction results for unmodeled PSF variations for NGC 2257 as an example. We find a star's 50 closest main-sequence neighbours (in position) and compute their median distance from the median ridge line. We then shift the star by that amount. Note the colourbar scale is not the same in the three panels.}\label{PSF}
\end{figure*}

Every CCD has unmodeled PSF variations that are correlated with position, which cause errors in the photometry of these stars \citet{Milone2012}. In addition, there is differential reddening effects. In order to deal with these two difficulties, stars that are close to each other on the CCD will exhibit similar PSF variations. In order to account for this, we calculate the median distance between the 50 closest main-sequence stars to a single main-sequence star and the median ridge line of the main sequence. If there were no PSF variations, then the median position away from the median ridge line would be zero; there would be an equal number of stars redder and bluer than the median ridge line. We repeat this for every star and its 50 closest main-sequence stars and shift that star by this correction factor. The results are shown in Figure \ref{PSF} for NGC 2257 in the three colours. The average of the corrections is $\sim\pm$0.02 mag for each of the colours. We reiterate that this step is only performed for main-sequence stars, not for all of the stars of each cluster.

\begin{figure*}
  \centering
  \begin{tabular}{cc}
    \includegraphics[width=\columnwidth]{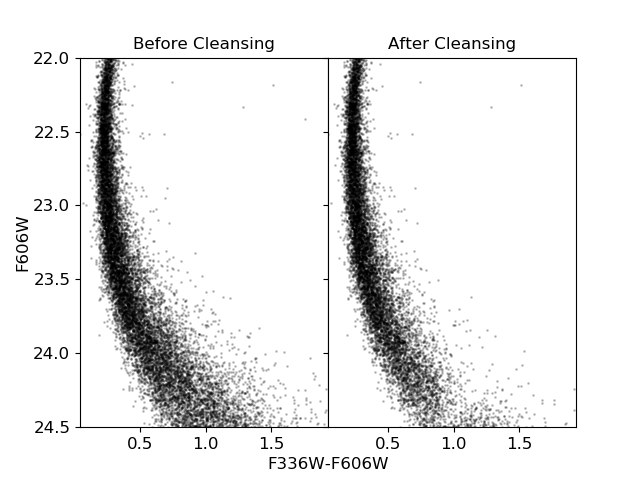} &
        \includegraphics[width=\columnwidth]{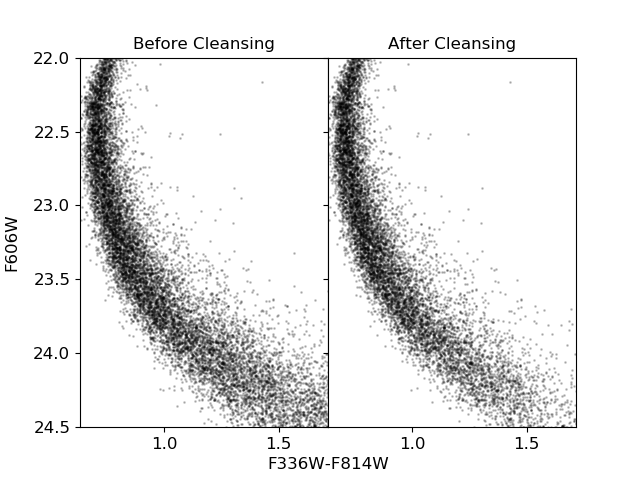}\\
        \small (a) &
                        \small (b) \\
  \end{tabular}
  \caption{The main-sequence of NGC 2257 before and after the cleansing process. We remove stars with large photometric errors and shift their colours as described in Section \ref{Data}. It is clear that the data cleansing creates a tighter and cleaner CMD.}\label{Cleansing}
\end{figure*}

Figure \ref{Cleansing} shows the main-sequence (MS) of NGC 1466 before and after the data cleansing process in two different colours, F336W-F606W and F336W-F814W. It is clear that the data become tighter and cleaner after the data cleansing process.

With this photometry, we create a straightened CMD based on the median ridge line of the MS. Straightening the CMD allows us to better view the multiple populations, if any, evident in the MS. To create our straightened CMDs, we first use the straightened histograms method presented in \citet{Marin-Franch}. We divide the stars into bins based on magnitude with widths of 0.1 mag to 0.3 mag determined by what creates the best fit. The bins overlap by 0.05 mag. We then find the median ridge line of each bin, which we use to straighten the CMD into a straightened colour-magnitude coordinate system.

To determine whether or not our straightened histograms show evidence of a redder population, we create histograms along the straightened colour-magnitude coordinate system in bins of 1.0 magnitude in the straightened colour-magnitude space. Changing the size of the magnitude bins does not greatly affect the results. We fit a double Gaussian equation: 

\begin{equation}
    f=(1-w)\exp\left(\frac{\left(x_{1}-\mu_{1}\right)^{2}}{2\sigma_{1}^{2}}\right)+w\exp\left(\frac{\left(x_{2}-\mu_{2}\right)^{2}}{2\sigma_{2}^{2}}\right)
    \label{Gauss}
\end{equation}

to the histogram using linear least squares. The $w$ parameter is the percentage that the redder population contributes to the overall population and we refer to it is as the weight of the redder population. However, there is some degeneracy between the different fitting parameters and therefore we caution that the exact value of the weight parameter is uncertain and should be seen as an estimate of the relative size of the redder population.

\section{Results}

\subsection{Main-Sequence}

\begin{figure*}
  \centering
  \begin{tabular}{cc}
    \includegraphics[width=\columnwidth]{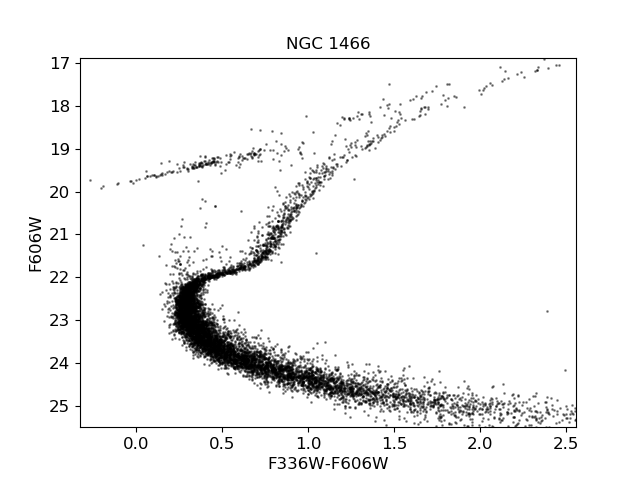} &
        \includegraphics[width=\columnwidth]{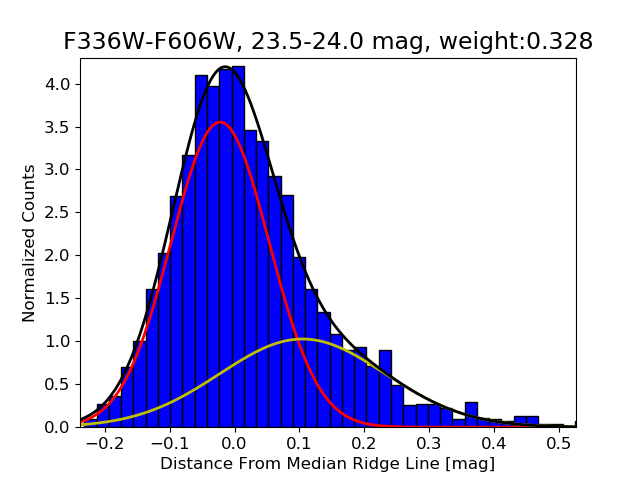}\\
        \small (a) &
                        \small (b) \\

  		\includegraphics[width=\columnwidth]{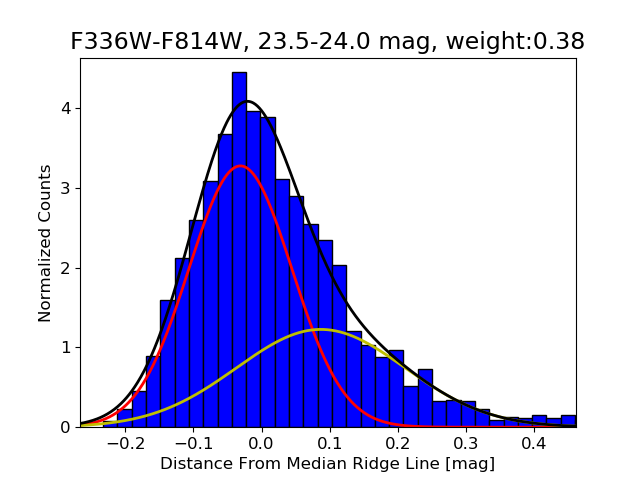} &
  		\includegraphics[width=\columnwidth]{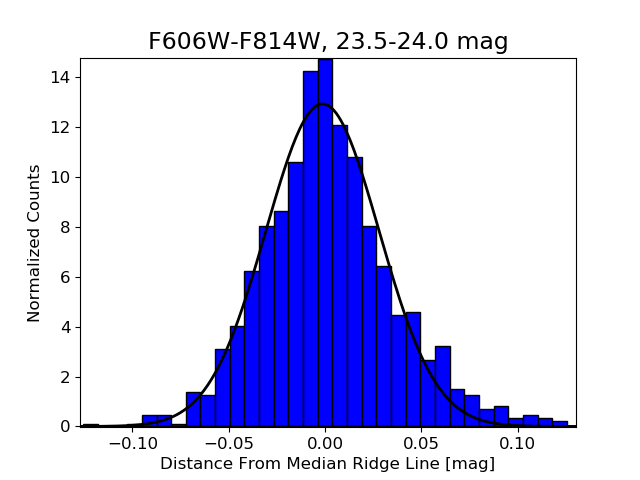}
		\\
                 \small (c) & \small (d)
  \end{tabular}
  \caption{Full CMD of NGC 1466 for the F336W-F606W colour (a). This CMD is a result of the data cleansing process. Panels (b), (c), and (d) are straightened histograms along the median ridge line. There is evidence for multiple populations in the F336W-F606W and F336W-F814W colours but not in the F606W-F814W colour as is expected. The weight of the redder population is between 30\% and 40\%. The distance between the centers of the two populations is $\sim 0.15$ mag.}
  \label{NGC1466}
\end{figure*}

\begin{figure*}
  \centering
  \begin{tabular}{cc}
    \includegraphics[width=\columnwidth]{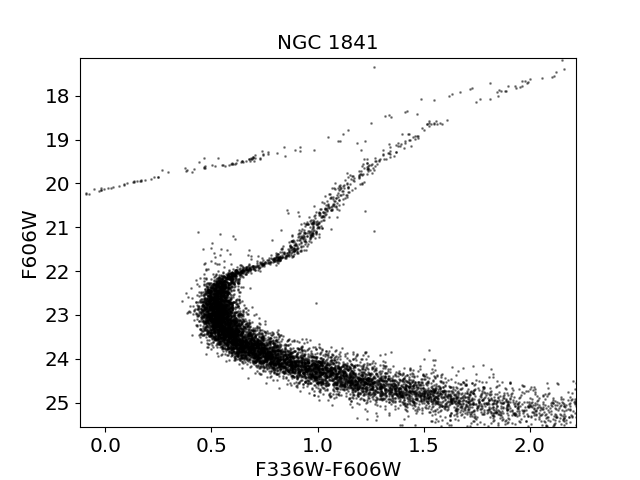} &
        \includegraphics[width=\columnwidth]{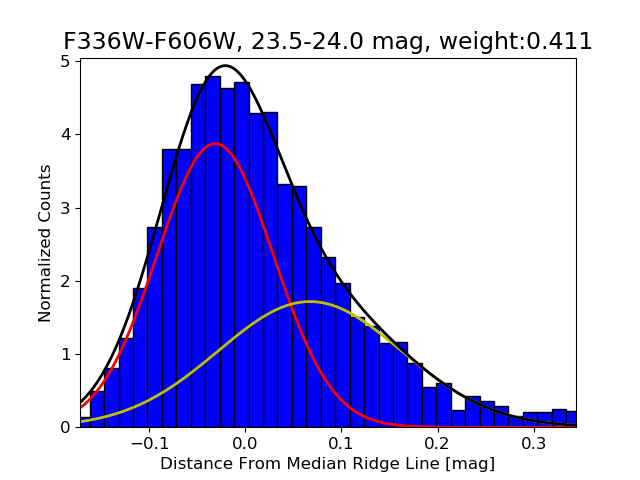}\\
        \small (a) &
                        \small (b) \\

  		\includegraphics[width=\columnwidth]{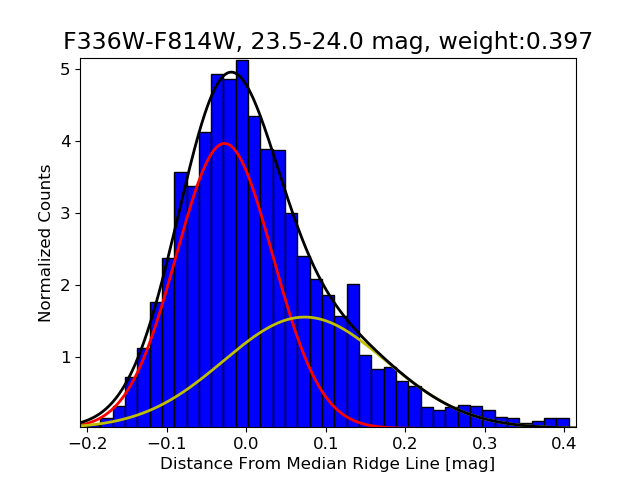} &
  		\includegraphics[width=\columnwidth]{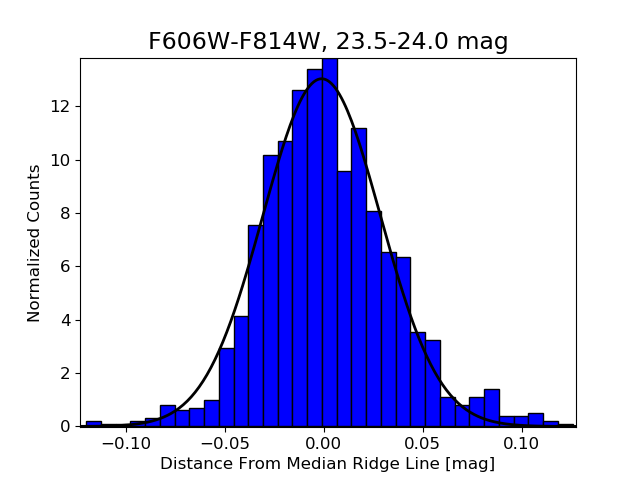}
		\\
                 \small (c) & \small (d)
  \end{tabular}
  \caption{Full CMD of NGC 1841 for the F336W-F606W colour (a). This CMD is a result of the data cleansing process. Panels (b), (c), and (d) are straightened histograms along the median ridge line. There is evidence for multiple populations in the F336W-F606W and F336W-F814W colours but not in the F606W-F814W colour as is expected. The weight of the redder population is $\sim$40\%. The distance between the centers of the two populations is $\sim 0.08$ mag.}\label{NGC1841}
\end{figure*}

\begin{figure*}
  \centering
  \begin{tabular}{cc}
    \includegraphics[width=\columnwidth]{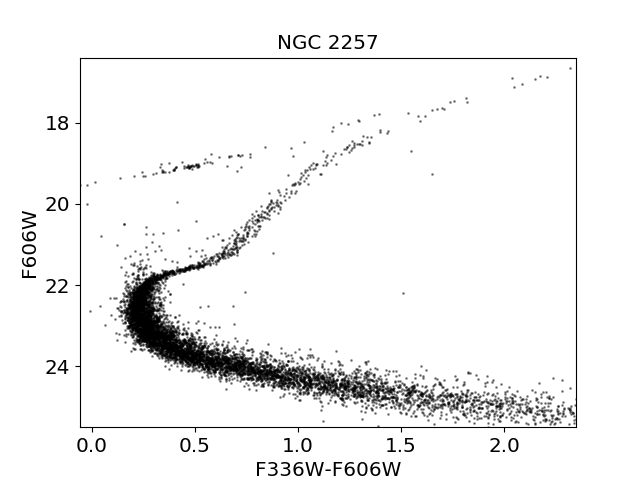} &
        \includegraphics[width=\columnwidth]{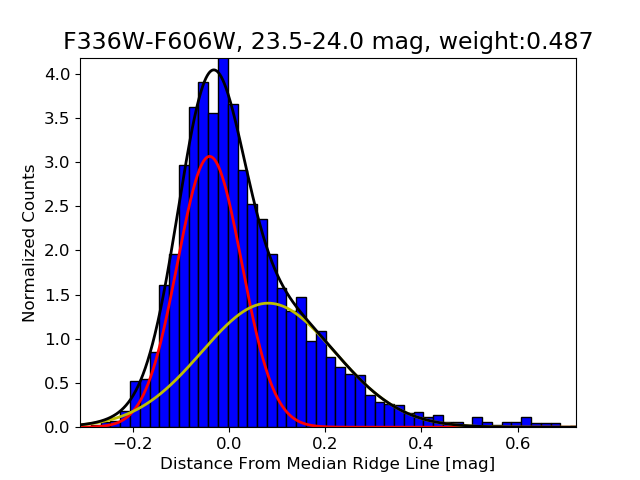}\\
        \small (a) &
                        \small (b) \\

  		\includegraphics[width=\columnwidth]{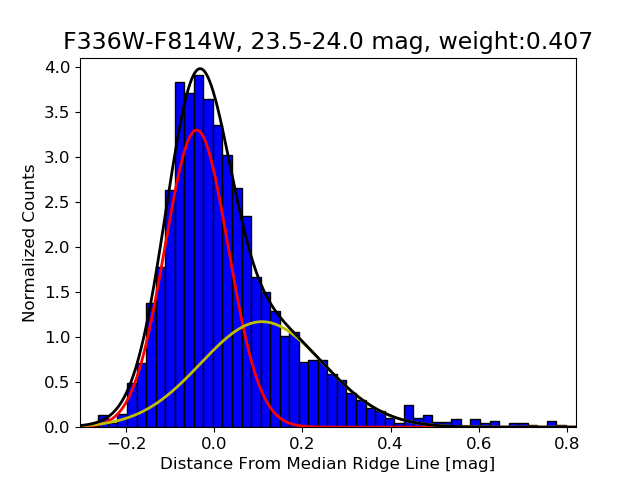} &
  		\includegraphics[width=\columnwidth]{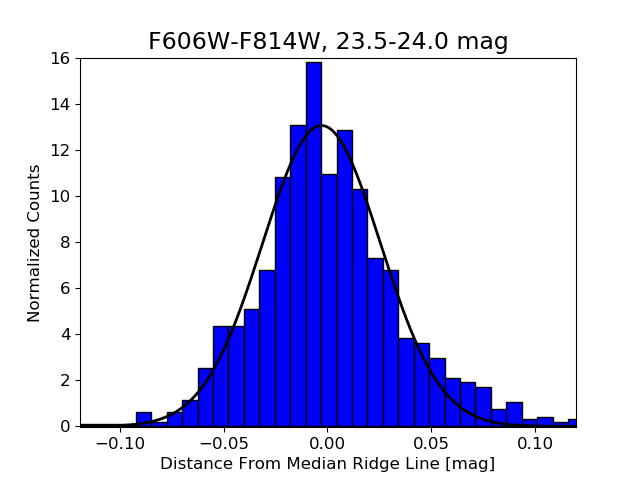}
		\\
                 \small (c) & \small (d)
  \end{tabular}
  \caption{Full CMD of NGC 2257 for the F336W-F606W colour (a). This CMD is a result of the data cleansing process. Panels (b), (c), and (d) are straightened histograms along the median ridge line. There is evidence for multiple populations in the F336W-F606W and F336W-F814W colours but not in the F606W-F814W colour as is expected. The weight of the redder population is between 40\% and 50\%. The distance between the centers of the two populations is $\sim 0.09$ mag.}\label{NGC2257}
\end{figure*}

\begin{figure*}
  \centering
  \begin{tabular}{cc}
    \includegraphics[width=\columnwidth]{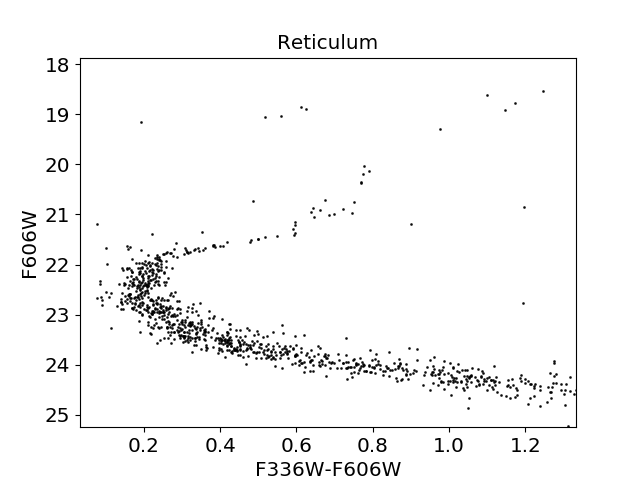} &
        \includegraphics[width=\columnwidth]{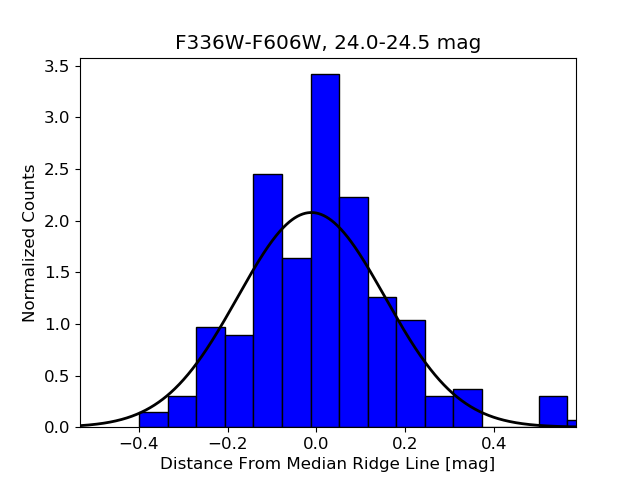}\\
        \small (a) &
                        \small (b) \\

  		\includegraphics[width=\columnwidth]{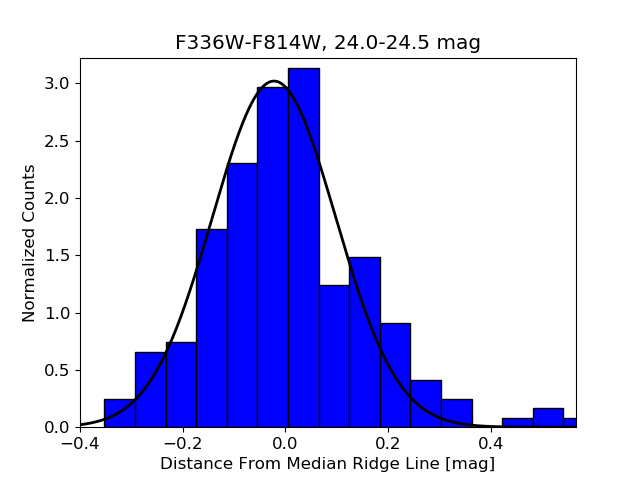} &
  		\includegraphics[width=\columnwidth]{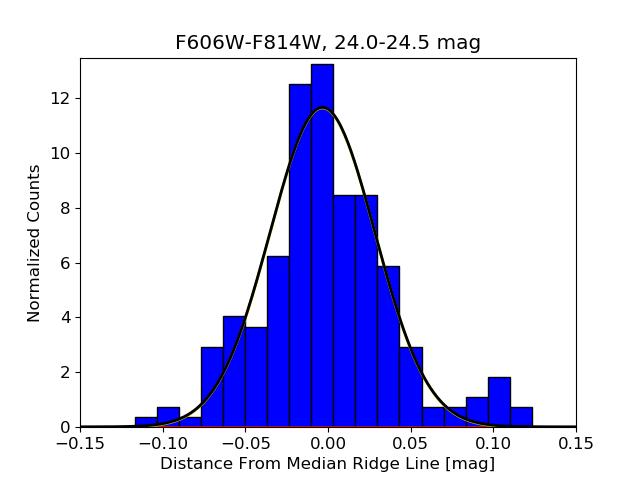}
		\\
                 \small (c) & \small (d)
  \end{tabular}
  \caption{Full CMD of Reticulum for the F336W-F606W colour (a). This CMD is a result of the data cleansing process. Panels (b), (c), and (d) are straightened histograms along the median ridge line. There is no evidence for multiple populations in Reticulum in any of the three colour combinations. As discussed in the text, this may be due to the small number of stars retrieved from our photometric analysis.}\label{Reticulum}
\end{figure*}

After performing the straightened histogram analysis on all four of the clusters, three out of our four clusters show evidence for a redder population along the main-sequence. Only in Reticulum does our analysis show preference for a single Gaussian population over a two Gaussian population. The histograms along with the F336W versus F336W-F606W CMD for NGC 1466, NGC 1841, NGC 2257, and Reticulum are shown in Figures \ref{NGC1466}, \ref{NGC1841}, \ref{NGC2257}, and \ref{Reticulum} respectively. The weight of the redder population in those three clusters is $\sim30-40\%$. However, there is some degeneracy in the precise value of this weight parameter so it is best to use it as an approximate size of the redder population with respect to the other population. The second population is evident along the CMD for around 3 mag fainter than the MSTO, with the relative size of the redder population varying on the order of $\pm$5\% depending on the magnitude range examined. The separation between the two populations becomes wider with increasing magnitude, but the photometric errors dominate the distribution for the faintest stars.

To test that the two populations are statistically different different from a single population, we use the two-population Kolmogorov-Smirnov (KS) test. Using a synthetic set of stars with the same luminosity function as each of the three clusters, we vary the multiple population parameters ($\mu$, $\sigma$, $x_{2}$, and $w$) and draw samples from this distribution. We find the best fitting parameters using the KS test to see which model population most closely matches the data. The parameters that we find through this method closely match the parameters derived from our double Gaussian fit. For the F606W-F814W we do not find any evidence for a split population using the KS test with a likelihood of greater than 99.99\%. 

\begin{table}
\caption{Population separations}
\label{separationTable}
\begin{tabular}{lcc}
\hline
Cluster & F336W-F606W & F336W-F814W\\
\hline
NGC 1466 & 0.28 & 0.15\\
NGC 1841 & 0.08 & 0.07\\
NGC 2257 & 0.10 & 0.13\\
\hline
\end{tabular}
\end{table}

We find a separation of the two populations is shown in Table \ref{separationTable}. The separation of the two populations roughly matches what has been seen in NGC 6397 \citep{NGC6397} and 47 Tuc \citet{Milone2012}. In \citet{NGC6397}, the separation of the two populations in NGC 6397 in F275W-F336W is $\sim$0.05 mag and in \citet{Milone2012} for 47 Tuc, the separation in F336W-F435W is $\sim$ 0.05 mag just below the MSTO and $\sim$ 0.2 mag farther below the MSTO. However, direct comparison is difficult due to the lack of overlap between our filters and the filters used in other studies. According to the bolometric corrections presented in \citet{Milone2012}, for the F606W-F814W there is very little separation in two populations that vary either in CNO abundances, He abundances, or a combination of both. For the F336W-F814W color, a difference in He abundance would cause a population separation of about 0.03 mag, which is less than we are seeing. A separation of about 0.1 mag is expected for CNO abundance differences. A combination of  CNO and He abundance variations would be expected to cause a separation of $\sim$ 0.06 mag. With this in mind, it is more likely that the main driver for the spread along the MS for NGC 1466 and NGC 2257 to be caused by CNO abundance variations than He abundance variations. For NGC 1841, it is more difficult to distinguish between the three possible scenarios.

Another possibility for the presence of a redder population of stars seen on our straightened CMD is due to binary star contamination. Binary stars appear redder in every colour since the total luminosity is increased due to the presence of two stars while the temperature remains roughly constant. To ensure that the observed second population is not due to binary contamination, we use a technique from \citet{Cummings2014}. We select all of the stars that are at least 0.035 mag redder in every colour (each star is redder in every colour combination), since binaries should also affect the F606W-F814W colour while the multiple populations have a negligible effect on this colour. In all of the clusters, the total number of these redder binaries in any magnitude bin was less than 10\% and therefore does not account for the entirety of the redder population. 

However, this technique for detecting binaries does not work at differentiating a single star from a binary with a small mass ratio. In order to determine if these small mass ratio binaries are a significant fraction of our observed redder population, we create a Monte Carlo simulation following a similar procedure to \citet{Binaries}. We simulate binaries using a flat distribution in colour up to 0.7 mag redder than the median ridge line. We incorporate these binaries into a set of straightened histograms following the luminosity function of each GC. We vary the total fraction of binary stars from 1\% up to 50\%. The resulting histograms' features do not match what we see in the histograms of the same magnitude range. The redder tail of the distribution is less extended and have smaller amplitudes in our simulations than what we observe in our GCs. Therefore, we can confidently say that our redder second population is not exclusively caused by binary contamination.

There is some evidence that the multiple populations in GCs are radially dependent. Some models that use rotating massive stars or AGB stars to create the multiple populations tend to create a radially dependent redder population that is less centrally located \citep{Larsen}. Some evidence of this has been seen for this effect in \citet{Milone2012}, \citet{Larsen}, and \citet{Dalessandro}. To test the radial location of our redder population, we create radial cuts based on distance from the centre of the cluster in pixel space. We use 500, 1000, and 2000 pixel radius cuts. Unlike NGC 2210 in \citet{Me}, applying each of the radial cuts to every cluster does not change the presence of the redder population for all four of the clusters. Varying the radial cut for each cluster affects the weight of the redder population on the order of $\pm$ 3\% which is within the variations expected. Thus, we do not find evidence for radial effects in the fraction of the redder population of stars.

We do not find evidence for a secondary population in Reticulum. However, we only retrieve $\sim$5500 stars from our photometry of Reticulum, which is in contrast to the other clusters where we retrieve $\sim$20,000 stars. The small number statistics could be hiding evidence for an extended, red population. To test this idea, we create many sample histograms that follow the redder population parameters of the other three clusters (their weight, standard deviations, and distance between peaks). We draw 5000 stars from this distribution and create a histogram. Next, we use our Gaussian fitting on this created sample and see whether the one or two population model is determined to be the better fit. In over 99\% of the cases, the one population model is the preferred model by our fitting routine. Even in the cases where the two population model was preferred, there is only a small difference in the goodness of fit of both the one and two population models, and therefore is not statistically significant. Therefore, it is likely that even if Reticulum had a split or broadened main-sequence, we would not be able to recover it in our analysis due to the limited sample size.

\subsection{Red Giant Branch}


Split RGBs have been seen in virtually all Galactic GCs observed with the `magic trio' of filters, as well as older, massive Galactic GCs observed with the Martocchia filters F336W, F343N, and F438W.

To examine the RGB, we follow a similar procedure as laid out for the MS. However, since there are fewer stars present in the RGB as compared to the MS, the stars are not colour corrected with respect to the RGB. Instead, we find the nearest MS neighbour in pixel-space and use that star's colour correction. However, the colour correction is not as dramatic as along the MS due to the RGB having fewer stars than the MS. We remove stars with higher intrinsic errors and straighten along the median ridge line. Since the relative lack of stars makes finding the median ridge line more difficult along the RGB, we also smooth the RGB's median ridge line using a polynomial fit. However, not including this smoothing does not greatly effect the histograms.

\begin{figure*}
  \centering
  \begin{tabular}{cc}
      \includegraphics[width=0.85\columnwidth]{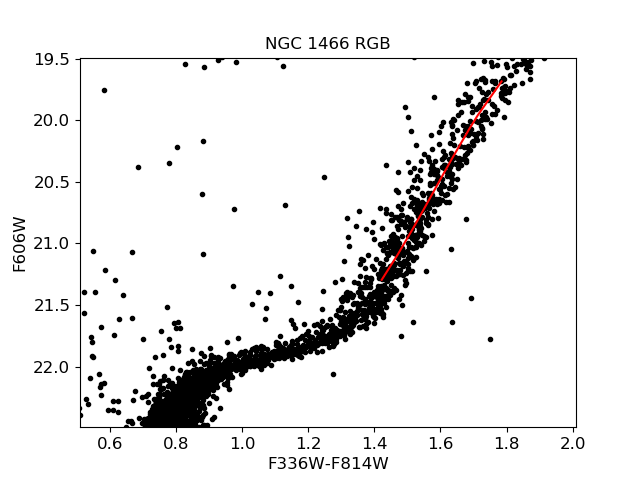} &
        \includegraphics[width=0.85\columnwidth]{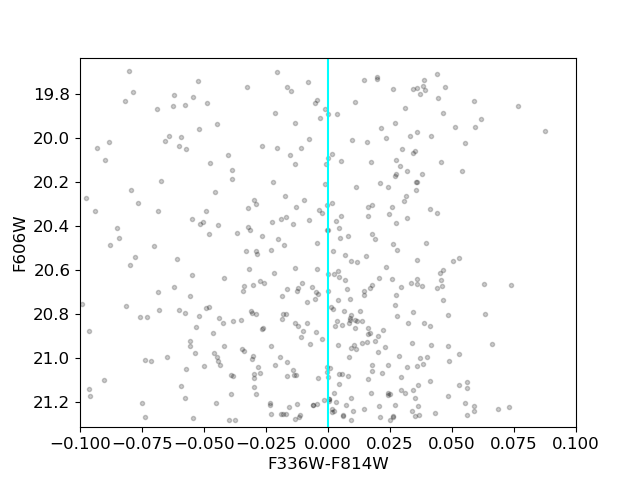}\\
        
                \small (a) &
                        \small (b) \\
    \includegraphics[width=0.85\columnwidth]{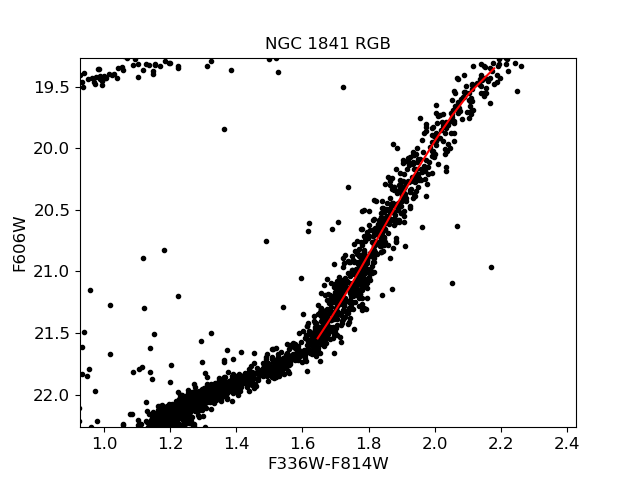} &
        \includegraphics[width=0.85\columnwidth]{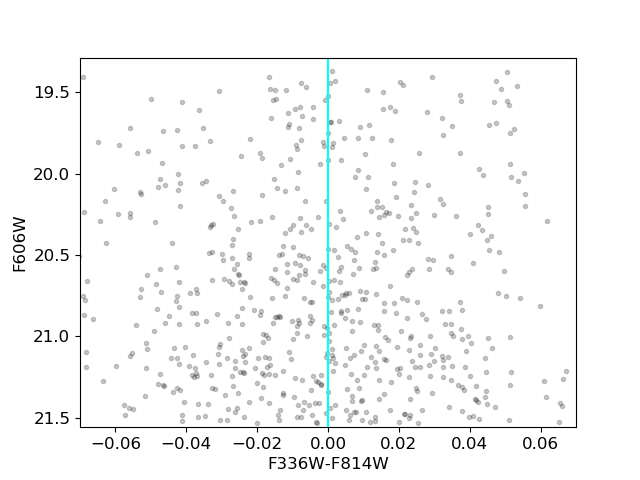}\\
        
                \small (c) &
                        \small (d) \\
    \includegraphics[width=0.85\columnwidth]{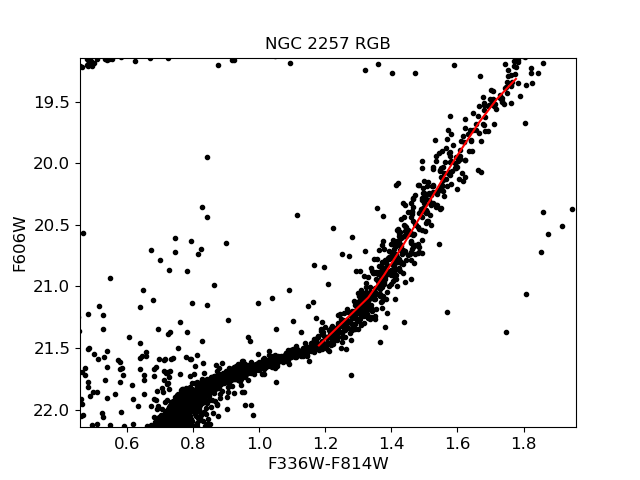} &
        \includegraphics[width=0.85\columnwidth]{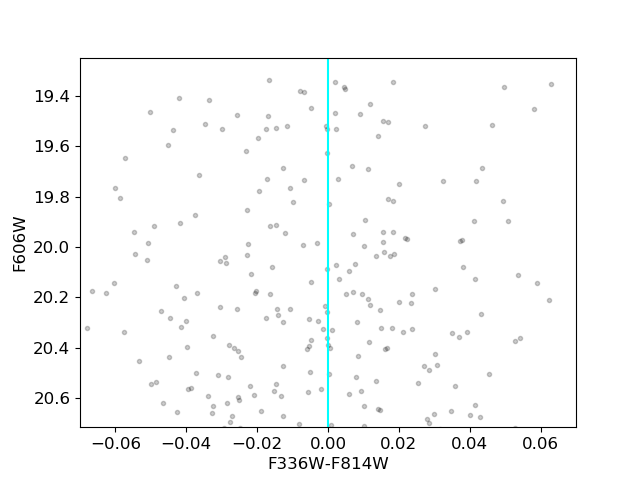}\\
        
                \small (e) &
                        \small (f) \\
    \includegraphics[width=0.85\columnwidth]{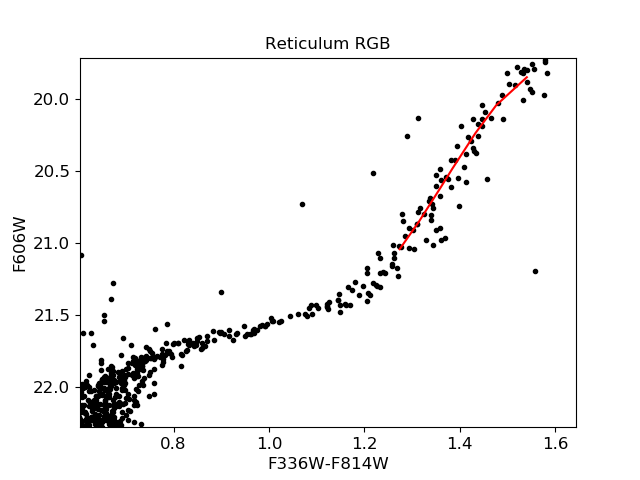} &
        \includegraphics[width=0.85\columnwidth]{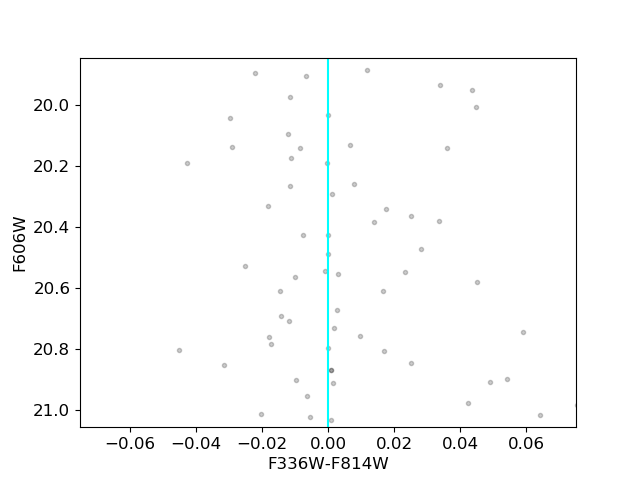}\\
        
                \small (g) &
                        \small (h) \\
        
  \end{tabular}
  \caption{Median ridge line along the RGB and straightened RGB for NGC 1466, NGC 1841, NGC 2257, and Reticulum respectively. The cyan line in the right panels shows the median ridge line. }\label{RGB}
\end{figure*}

Figure \ref{RGB} shows the CMDs for each of our GCs for the F336W-F814W colour along with the straightened histogram for the same colour. NGC 1841, NGC 2257, and Reticulum do not show evidence for multiple populations in the RGB in any colour. NGC 1466 seems to have a wide RGB in the F336W-F606W and F336W-F814W colours, but a single Gaussian fit is the preferred solution as opposed to a two Gaussian solution. We perform a Monte Carlo simulation similar to what we perform for the MS of Reticulum to test whether the lack of a redder population is due to small number statistics. We create a simulated distribution of stars using a similar distribution as NGC 1978 from \citet{Martocchia} but a luminosity distribution matching NGC 1466 along the RGB. Only when the number of stars is increased from $\sim$ 500 (the number that are measured) to $\sim$ 1500 stars, we start to see a preference for the two population Gaussian model instead of the single population Gaussian model. Consequently, even if there was a similar redder population in NGC 1466 as there is in NGC 1978, we would not be able to retrieve it with our data. As an additional test, we perform a similar Monte Carlo simulation but this time separate the two populations by 0.15 mag along the RGB. This width is similar to what is found on the MS. In addition, we convolve this separation with the observational errors of the RGB stars. We find that the observational errors are large enough to smear the separations of two populations along the RGB. The width for the F336W-F606W recovered from our simulation is 0.19 mag while we measure a width of 0.17 mag for the same F606W magnitude. For the F336W-F814W color, the simulated width is 0.2 mag while the measured width is 0.18 mag. For the F606W-F814W color, the simulated width is 0.05 while the observed width is 0.06 mag. There are a few ($\sim10$) RGB stars that are outside this width but they are not numerous enough for our program to prefer a two-population Gaussian fit over a single population Gaussian. Therefore, we do not find evidence for two populations along the RGB but this could be due to observational errors and the smaller sample of RGB stars we have compared to the MS.

NGC 2257 also appears to have a wide RGB. The most likely reason that we do not see evidence for split RGBs in these old, massive clusters, while they have been seen in similar, old, massive clusters (e.g. \citet{Milone2012} and \citet{Martocchia}) is that we do not employ filters that are as sensitive to the light element abundance variations as used in these other studies.

\subsection{Horizontal Branch}

\begin{figure*}
  \centering
  \begin{tabular}{cc}
\includegraphics[width=0.85\columnwidth]{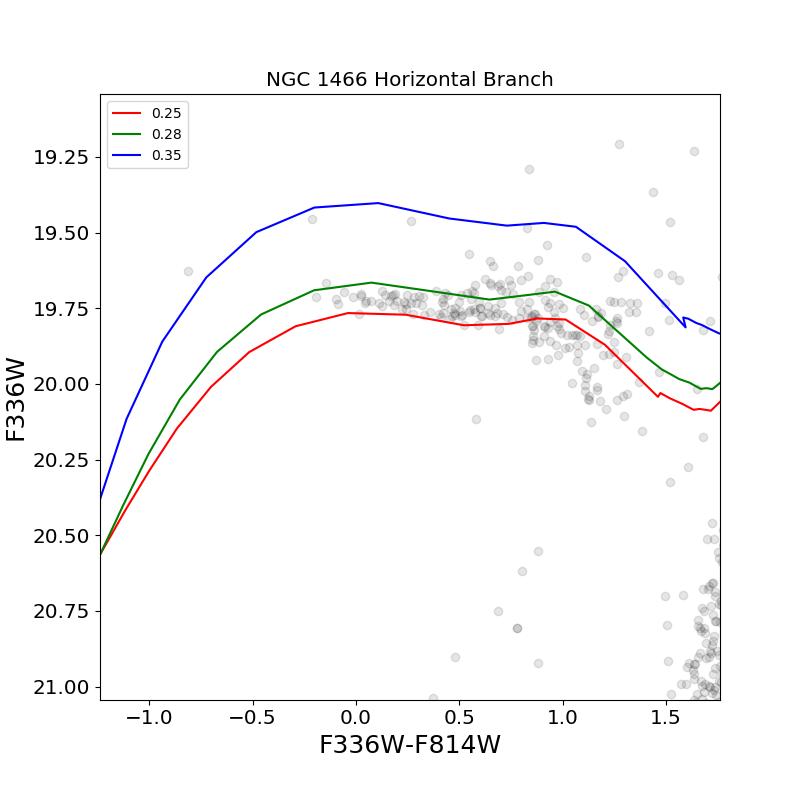} &
\includegraphics[width=0.85\columnwidth]{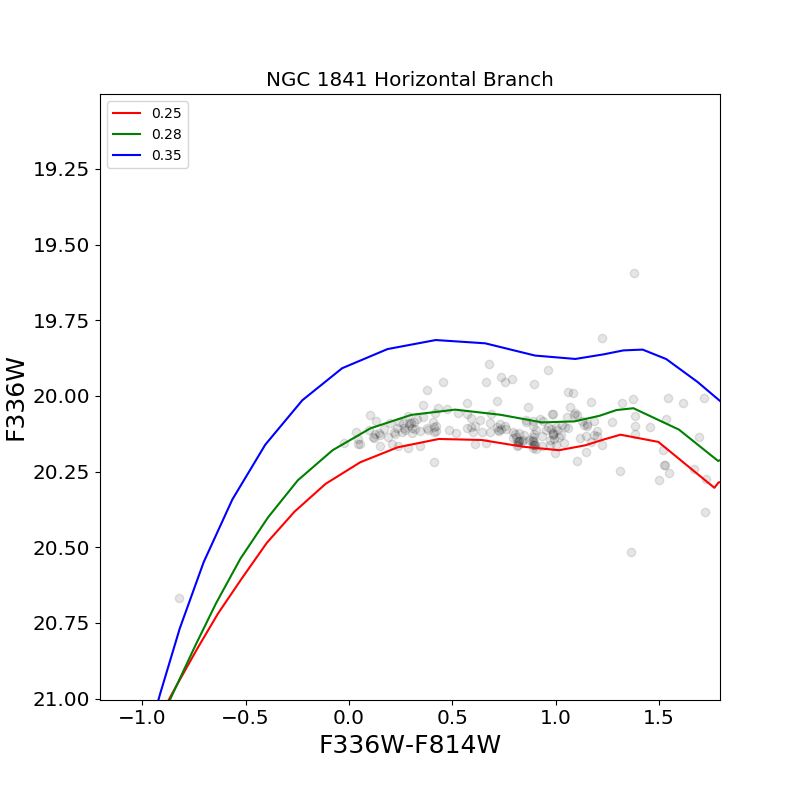}\\
                \small (a) &
                        \small (b) \\
\includegraphics[width=0.85\columnwidth]{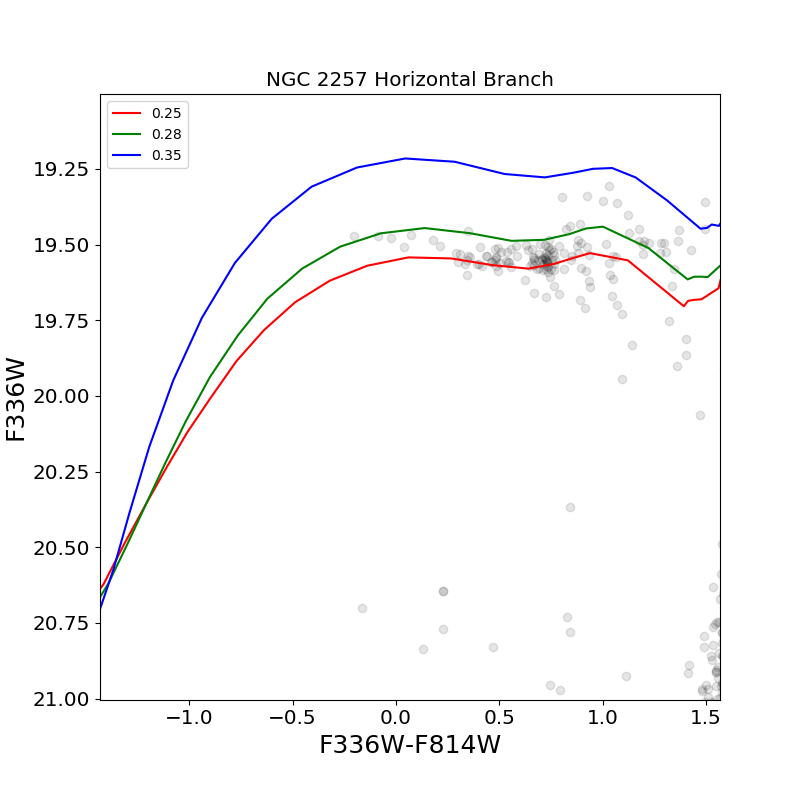} &
\includegraphics[width=0.85\columnwidth]{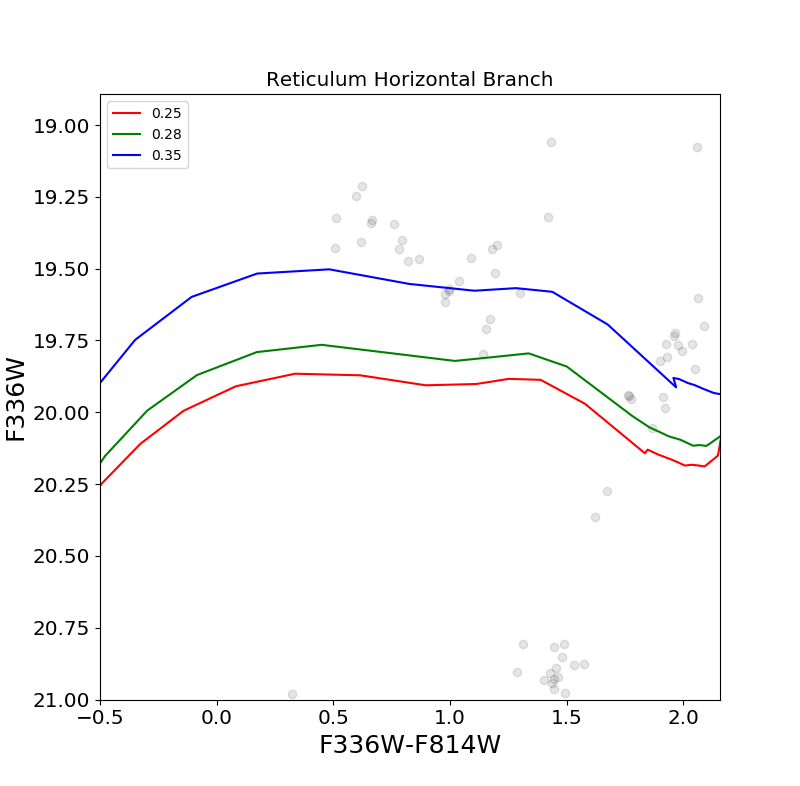}\\
                \small (c) &
                        \small (d) \\
\end{tabular}
\caption{Horizontal branches of NGC 1466, NGC 1841, NGC 2257, Reticulum respectively in the F336W-F814W colour. Overlaid are HB tracks from \citet{Basti} with varying Helium content. Unlike NGC 2210 in \citet{Me}, none of these clusters show evidence for multiple populations in the HB.}
\label{HB}
\end{figure*}

In \citet{Me}, we found that Hodge 11 has evidence for multiple populations along the horizontal branch (HB) with two fairly distinct clumps of stars seen in the CMDs of that cluster. Figure \ref{HB} shows each GC's HB in the F606W versus F336W-F814W plane. The theoretical HB tracks vary in total helium content \citep{Basti}. We do not see any evidence for large helium enhancement along the HB but we cannot fully rule out the presence of multiple populations along the HB. 

        
        


\section{Conclusion}

Three out of the four LMC GCs we analyzed in this work have strong evidence for multiple populations along the MS. This brings the total of ancient LMC GCs with multiple populations to 5. The only ancient LMC GC that we examined that does not show evidence for a redder population, Reticulum, is affected by small number statistics, which could be hiding the presence of a redder population. Its mass ($1.4\times 10^{5}M_{\odot}$ \citet{Piatti}) is above the limit of around a few $\times10^{4}M_{\odot}$ below which some clusters are found which do not harbor multiple populations (e.g. \citet{MartocchiaNew}). The abundance variations that cause these redder populations cannot be known without further spectroscopic studies or further photometric studies with the `magic trio' of filters.

\vspace{3mm}

This research has made use of NASA's Astrophysics Data System. Based on observations made with the NASA/ESA Hubble Space Telescope, obtained from the data archive at the Space Telescope Science Institute. STScI is operated by the Association of Universities for Research in Astronomy, Inc. under NASA contract NAS 5-26555. This work was supported in part by STScI through a grant HST-GO-14164. DG gratefully acknowledges support from the Chilean Centro de Excelencia en Astrof\' isica y Tecnolog\' ias Afines (CATA) BASAL grant AFB-170002. DG also acknowledges financial support from the Direcci\' on de Investigaci\' on y Desarrollo de la Universidad de La Serena through the Programa de Incentivo a la Investigaci\' on de Acad\' emicos (PIA-DIDULS). DM gratefully acknowledges support from an Australian Research Council (ARC) Future Fellowship (FT160100206). SV gratefully acknowledges the support provided by Fondecyt reg. n. 1170518.




\bibliographystyle{mnras}
\bibliography{LMC} 





\bsp	
\label{lastpage}
\end{document}